\documentclass[namedreferences]{solarphysics}
\usepackage[hyperref,optionalrh,solaromanenum]{spr-sola-addons} 
\usepackage{graphicx}                    
\usepackage{color}                       
\usepackage{rotating}
\usepackage{soul}
\usepackage{breakurl}                         

\begin{document}

\begin{article}

\begin{opening}

\title{Empirical Model of 10\,--\,130\,MeV Solar Energetic Particle Spectra at 1\,AU Based on Coronal Mass Ejection Speed and Direction}

%
\author[addressref={aff1,aff2},email={alessandro.bruno-1@nasa.gov}]{\inits{A.}\fnm{Alessandro}~\lnm{Bruno}\orcid{0000-0001-5191-1662}}
\author[addressref={aff1,aff3},email={ian.g.richardson@nasa.gov}]{\inits{I.G.}\fnm{Ian~G.}~\lnm{Richardson}\orcid{0000-0002-3855-3634}}
%
\runningtitle{Empirical Model of 10\,--\,130\,MeV SEP Spectra at 1\,AU Based on CME Speed and Direction}
\runningauthor{A. Bruno \& I.G. Richardson, doi:10.1007/s11207-021-01779-4}

\address[id=aff1]{Heliophysics Division, NASA Goddard Space Flight Center, Greenbelt, MD, USA}
\address[id=aff2]{Department of Physics, Catholic University of America, Washington DC, USA}
\address[id=aff3]{Department of Astronomy, University of Maryland, College Park, MD, USA}

\begin{abstract}
We present a new empirical model to predict solar energetic particle (SEP) event-integrated and peak intensity spectra between 10 and 130\,MeV at 1\,AU, based on multi-point spacecraft measurements from the \textit{Solar TErrestrial RElations Observatory} (STEREO), the \textit{Geostationary Operational Environmental Satellites} (GOES), and the \textit{Payload for Antimatter Matter Exploration and Light-nuclei Astrophysics} (PAMELA) satellite experiment. The analyzed data sample includes 32 SEP events occurring between 2010 and 2014, with a statistically significant proton signal at energies in excess of a few tens of MeV, unambiguously recorded at three spacecraft locations. The spatial distributions of SEP intensities are reconstructed by assuming an energy-dependent 2D Gaussian functional form, and accounting for the correlation between the intensity and the speed of the parent coronal mass ejection (CME), and the magnetic-field-line connection angle. The CME measurements used are from the Space Weather Database Of Notifications, Knowledge, Information (DONKI). The model performance, including its extrapolations to lower/higher energies, is tested by comparing with the spectra of 20 SEP events not used to derive the model parameters. Despite the simplicity of the model, the observed and predicted event-integrated and peak intensities at Earth and at the STEREO spacecraft for these events show remarkable agreement, both in the spectral shapes and their absolute values.\\
\end{abstract}

%

\end{opening}

%
\section{Introduction}\label{Introduction}
Solar energetic particle (SEP) events are major space-weather disturbances both in the heliosphere and in the near-Earth environment, which significantly constrain human activities in space by posing serious radiation hazards for satellites, avionics, astronauts, and aircraft passengers on polar routes (e.g. \citealp{ref:SHEASMART2012}). Modern society's vulnerability to space-weather effects is exacerbated by its increasing reliance on technological systems (e.g. \citealp{ref:EASTWOOD2017}). Predicting SEP event occurrence and impact is therefore of crucial importance, especially in view of planned long-duration missions to the Moon and to Mars, beyond the Earth's protective atmosphere and magnetosphere (e.g. \citealp{ref:CUCINOTTA2010}).

Large SEP events of space-weather interest are believed to be primarily accelerated by coronal mass ejection (CME)-driven shocks (e.g. \citealp{ref:DESAIGIACALONE2016}). The acceleration efficiency is predicted to depend on several concomitant factors, including the shock speed, geometry, and age (e.g. \citealp{ref:TYLKA2005,ref:ZANK2007}), the coronal magnetic-field strength and configuration (e.g. \citealp{ref:KONG2017,ref:KONG2019}), 
the presence of seed-particle populations (e.g. \citealp{ref:KAHLER2000,ref:KAHLER2001,ref:TYLKA2005,ref:CLIVER2006,ref:DESAI2006,ref:DESAI2016})
and pre-existing turbulence (e.g. \citealp{ref:GOPALSWAMY2004,ref:LI2012,ref:DING2013,ref:ZHAO2014}). 
Although they might be included in a full physics-based SEP prediction model, most of these factors are difficult to evaluate or are not currently measurable directly. A number of authors have reported a significant correlation between SEP event intensities and the parent CME speeds, which has been interpreted as evidence that SEPs are accelerated by shocks (e.g. \citealp{ref:CANE2010,ref:GOPALSWAMY2002,ref:GOPALSWAMY2004,ref:KAHLER1978,ref:KAHLER1984,ref:KAHLER1987,ref:KAHLERVOURLIDAS2005,ref:LARIOKARELITZ2014,ref:REAMES2000,ref:RICHARDSON2014,ref:RICHARDSON2015}). In particular, the most energetic SEP events, such as those causing the so-called ground-level enhancements (GLEs), were found to be typically associated with the fastest ($\approx$2000\,km s$^{-1}$ average velocity) CMEs \citep{ref:GOPALSWAMY2006}.
Such observations suggest that the CME velocity might be used as a proxy for the shock-acceleration efficiency in an empirical SEP prediction model.

Another important factor determining the relative particle intensity during an SEP event -- as well as the occurrence of GLEs -- is the connection angle, defined as the angular distance between the SEP source at the Sun and the footpoint of the interplanetary magnetic-field (IMF) line passing the observing spacecraft. As a number of studies have shown, measured particle intensities tend to decrease with increasing connection angle (e.g. \citealp{ref:VANHOLLEBEKE1975,ref:KALLENRODE1993,ref:LARIO2006,ref:LARIO2013,ref:RICHARDSON2014,ref:RICHARDSON2017,ref:GOPALSWAMY2014}). 
In addition, \citet{ref:CANE1988} demonstrated that the longitudinal distribution of SEPs and its energy dependence, is also determined by the connection to the CME-driven shock. 
Since its launch in 2006, the \textit{Solar TErrestrial RElations Observatory} (STEREO) mission \citep{ref:KAISER2008} has enabled, in combination with near-Earth measurements, a multi-point investigation of SEP events over a wide longitudinal range with minimal radial gradient effects (e.g. \citealp{ref:LARIO2013,ref:RICHARDSON2014,ref:RICHARDSON2017,ref:COHEN2017}). 
However, the limited number of observation points have precluded any precise measurement of the longitudinal distribution, which has been usually assumed to be Gaussian in connection angle following \citet{ref:LARIO2006}.

\citet{ref:RICHARDSON2014} discussed the properties of all the $>$25\,MeV proton events observed by the STEREO spacecraft and/or at Earth during 2006\,--\,2013 and summarized the intensity of 14\,--\,24\,MeV protons in 25 SEP three-spacecraft events as 
\begin{equation}\label{eq:RICHARDSON2014}
\Phi(\beta)\approx 0.013 \hspace{1mm} \exp\left(0.0036\hspace{1mm} V_{\mathrm{cme}} - \beta^{2}/2\sigma_{\mathrm{sep}}^{2}\right),
\end{equation}
where $\Phi$ [in MeV$^{-1}$sr$^{-1}$s$^{-1}$cm$^{-2}$] is the peak intensity of the Gaussian fit to the event, $V_{\mathrm{cme}}$ [in km\,s$^{-1}$] is the speed of the related CME, $\beta$ is the longitudinal connection angle and $\sigma_{\mathrm{sep}}$ = 43$^\circ$ is the mean Gaussian standard deviation for these events. 
\citet{ref:RICHARDSON2018} used the relationship in Equation \ref{eq:RICHARDSON2014} to develop an empirical model (subsequently named ``SEPSTER'' -- SEP predictions based on STEREO observations) to predict the 14\,--\,24\,MeV proton peak intensity based on CME speed and direction relative to the observing spacecraft. 
The spatial width value [$\sigma_{\mathrm{sep}}$] derived by \citet{ref:RICHARDSON2014} is consistent with the results obtained by \citet{ref:LARIO2013} 
and \citet{ref:PAASSILTA2018} using the 25\,--\,53\,MeV and the $>$55\,MeV proton peak intensities, respectively. \citet{ref:COHEN2017} 
analyzed the 0.3, 1, and 10\,MeV\,n$^{-1}$ event-integrated intensities of H, He, O, and Fe ions, reporting a mean $\sigma_{\mathrm{sep}}$ value decreasing 
from $\approx$52$^\circ$ to $\approx$36$^\circ$ with increasing energy,  which they suggested is in line with theoretical expectations that higher-energy particles are 
efficiently accelerated over a smaller shock region around the ``nose'' or for shorter times as the shock expands (e.g. \citealp{ref:LEE2005,ref:CAPRIOLI2014,ref:DALLA2017}), or are less affected by field-line co-rotation effects (e.g. \citealp{ref:GIACALONE2012}).

Although typically neglected when compared to longitude-related effects, the SEP intensities at a given location are also affected by the latitudinal magnetic connectivity to the source \citep{ref:DALLA2010,ref:GOPALSWAMYMAKELA2014}. 
The latitude influence might be expected to become more important for higher-energy particles if they are accelerated by the strongest regions of a shock close to the nose or, in an alternative scenario, by the associated solar flare. Consistent with this prediction, \citet{ref:GOPALSWAMY2013,ref:GOPALSWAMY2014} showed that the latitudinal distance from the Ecliptic is typically larger for energetic eruptions not associated with GLE events, suggesting that the poorer connection to Earth was responsible. Although a direct estimate of the SEP event latitudinal spread was previously investigated in the pre-STEREO era by the \textit{Ulysses} mission (see, e.g., \citealp{ref:STRUMINSKY2006} and references therein), no similar observations are currently available.

Recently, \citet{ref:DENOLFO2019} derived the $>$80\,MeV proton spatial distribution of 14 events by means of a 2D Gaussian modeling accounting for both longitudinal and latitudinal connectivity, based on the assumption of a spherical symmetry of particle intensities. In the present work, following \citet{ref:RICHARDSON2014} and \citet{ref:RICHARDSON2018}, we exploit the widely reported correlation between particle intensities and CME speeds, as well as the spatial distribution reconstructed for a number of SEP events with the 2D approach of \citet{ref:DENOLFO2019}, to develop an empirical model that predicts 10\,--\,130\,MeV proton spectra at 1\,AU. The article is structured as follows: in Section \ref{Data analysis} we present the SEP events used to train the model, along with the spectral analysis of measured intensities. In Sections \ref{SEP intensity spatial distribution} and \ref{Dependence on CME speed} we describe the method for the reconstruction of the SEP spatial distribution, and the algorithm based on the CME speed and direction. In Sections \ref{Model uncertainties} and \ref{Model testing} we discuss the model uncertainties and test its performance. Finally, Section \ref{Summary and Conclusions} summarizes the study and presents our conclusions.

\section{Data Analysis}\label{Data analysis}
\subsection{Data Sample}
The starting point for developing the empirical SEP prediction formula is a set of 32 SEP events occurring between 2010 and 2014, in which proton intensities above a few tens of MeV were unambiguously measured at three 1\,AU locations by the twin STEREO spacecraft and near-Earth assets, including the \textit{Geostationary Operational Environmental Satellites} (GOES) and the \textit{Antimatter Matter Exploration and Light-nuclei Astrophysics} (PAMELA) experiment onboard the \textit{Resurs-DK1} Russian satellite (e.g. \citealp{ref:ADRIANI2017}). The selected period corresponds to the interval when all of these spacecraft were operating and numerous SEP events were detected at Earth and the STEREO locations (e.g. \citealp{ref:RICHARDSON2014,ref:PAASSILTA2018}), around the maximum of Solar Cycle 24.

\begin{table}[!th]
\center
\setlength{\tabcolsep}{4.9pt}
\begin{tabular}{ccrrcccc}
 & \multicolumn{4}{c}{CME} & \multicolumn{3}{c}{Spacecraft footpoints}\\
\# & Onset time & Speed & Width & Direction & STB & Earth & STA\\
\hline
1 & 2010-06-12T? & ... & ... & ... & S06E07 & N00W58 & N07W116 \\ 
2 & 2010-08-14T10:12 & 950 & 80 & N11W58 & N00W00 & N06W60 & N04W141 \\ 
3 & 2010-08-18T06:00 & 1091 & 80 & S30W97 & N00W08 & N06W63 & N03W139 \\ 
4 & 2010-09-08T23:26 & 850 & 90 & N08W96 & N02W00 & N07W55 & N00W133 \\ 
5* & 2011-03-21T02:54 & 1000 & 140 & N20W130 & S00E29 & S07W67 & N01W139 \\ 
6 & 2011-08-04T04:10 & 1950 & 120 & N14W40 & S03E45 & N05W64 & N02W154 \\ 
7* & 2011-09-06T22:40 & 650 & 60 & N20W20 & S00E43 & N07W57 & S01W174 \\ 
8 & 2011-09-22T11:24 & 1000 & 140 & N10E90 & N01E39 & N07W62 & S03W161 \\ 
9 & 2011-10-22T11:20 & 990 & 110 & N52W90 & N04E37 & N05W72 & S06E178 \\ 
10* & 2011-11-03T22:39 & 1100 & 130 & S07E160 & N05E22 & N04W64 & S06E173 \\ 
11 & 2011-11-26T07:12 & 930 & 144 & N23W45 & N06E46 & N01W57 & S07W160 \\ 
12* & 2012-01-23T04:00 & 2211 & 124 & N41W26 & N06E46 & S05W58 & S02W173 \\ 
13* & 2012-03-07T00:36 & 2200 & 100 & N30E60 & N03E53 & S07W62 & N02W169 \\ 
14* & 2012-05-17T01:48 & 1500 & 90 & S10W75 & S04E58 & S02W62 & N07W162 \\ 
15 & 2012-05-26T22:54 & 1100 & 140 & N05W110 & S05E81 & S01W58 & N07W173 \\ 
16* & 2012-07-23T02:36 & 3435 & 160 & S15W144 & S06E44 & N05W50 & N01W163 \\ 
17 & 2012-08-31T20:36 & 1498 & 150 & S15E63 & S03E37 & N07W63 & S03E178 \\ 
18 & 2013-02-06T00:36 & 1226 & 76 & N30E35 & N07E49 & S06W66 & N01E153 \\ 
19* & 2013-04-11T07:36 & 675 & 116 & N00E15 & N02E66 & S05W66 & N07W177 \\ 
20 & 2013-04-24T? & ... & ... & ... & N00E78 & S04W70 & N07W177 \\ 
21 & 2013-05-13T16:18 & 1900 & 80 & N10E70 & S01E81 & S01W60 & N03W175 \\ 
22* & 2013-05-22T13:24 & 1200 & 120 & N10W80 & S02E92 & S00W47 & N04E170 \\ 
23 & 2013-06-21T03:24 & 1970 & 140 & S19E57 & S05E100 & N00W44 & N04E172 \\ 
24 & 2013-08-19T23:12 & 1200  &  90 & N17W180 & S08E72 & N02W48 & N02E165 \\ 
25 & 2013-10-25T15:12 & 980 & 80 & N15E63 & S03E60 & N03W67 & S02E173 \\ 
26* & 2013-11-02T04:48 & 1078 & 150 & N05W145 & S00E90 & N02W62 & S02E162 \\ 
27 & 2013-11-19T10:39 & 910 & 100 & S29W82 & N01E78 & N01W54 & S03E143 \\ 
28 & 2013-12-26T03:40 & 1600 & 180 & S31E134 & N05E92 & S01W75 & S03E160 \\ 
29* & 2014-01-06T08:09 & 1275 & 90 & S03W102 & N05E92 & S01W51 & S02E150 \\ 
30* & 2014-01-07T18:24 & 2061 & 98 & S24W30 & N06E77 & S01W52 & S02E152 \\ 
31* & 2014-02-25T01:25 & 1670 & 132 & S11E78 & N03E119 & S02W48 & S00E149 \\ 
32* & 2014-09-01T11:24 & 1700 & 92 & N01E155 & S05E106 & N02W46 & S01E138 \\ 
\hline
\end{tabular}
*Associated with SEP events observed by PAMELA \citep{ref:BRUNO2018}.
\caption{CMEs associated with the SEP events analyzed in this work. The first column is the event number. Columns 2\,--\,5 report the CME first appearance time [UT], space speed [km s$^{-1}$], angular width [$^\circ$] and direction from the DONKI catalog. 
The CME parameters are not available for events \#1 and \#20.
The right three columns list the location of the footpoints of the Parker spiral field lines crossed by each spacecraft (STEREO-A/B and GOES/PAMELA) at CME onset, mapped ballistically back to 2.5\,R$_{\odot}$. Footpoints locations and CME directions are expressed in terms of HGS latitudes/longitudes.}
\label{tab:EventList}
\end{table}

The relevant parameters associated with the parent eruptions are reported in Table \ref{tab:EventList}.
The first column gives the event number. Columns 2\,--\,5 list the CME first appearance time [UT], space (3D) speed, width and direction in Stonyhurst heliographic (HGS) coordinates from the Space Weather Database Of Notifications, Knowledge, Information (DONKI) developed at the Community Coordinated Modeling Center (CCMC). These are based on the triangulation of the STEREO and the \textit{SOlar and Heliospheric Observatory} (SOHO) coronagraph measurements at a $\approx$21.5 solar radii (R$_{\odot}$) height \citep{ref:LIU2010,ref:MAYS2015}. The CME parameters are not available for events \#1 and \#20; the resulting sample includes CMEs with speeds ranging from 650 to 3454\,km s$^{-1}$. We choose to use CME parameters from DONKI because, in addition to being based on multi-point coronagraph observations, if available, these parameters include the CME direction which is not provided by most other CME catalogs. Furthermore, DONKI includes reports of observations of space-weather phenomena and their interpretation in real-time provided by the CCMC space-weather team, and hence it simulates how a CME-based SEP prediction scheme might be applied in a forecasting environment.
Finally, columns 6\,--\,8 show the HGS coordinates of the magnetic footpoints of the IMF lines passing through STEREO-B, the Earth, and STEREO-A, estimated at CME first appearance time and at a 2.5\,R$_{\odot}$ radial distance -- the nominal ``source surface'' height -- based on a simple IMF spiral model, as described in Appendix \ref{Appendix A}.

\subsection{SEP Spectral Analysis}\label{SEP spectral analysis}
Our multi-spacecraft analysis of SEP events is based on the proton intensities measured by the \textit{Solar Electron and Proton Telescope} (SEPT; \citealp{ref:MULLER2008}), the \textit{Low Energy Telescope} (LET; \citealp{ref:MEWALDT2008}), and the \textit{High Energy Telescope} (HET; \citealp{ref:VONROSENVINGE2008}) onboard the twin STEREOs, and by the \textit{Energetic Proton, Electron, and Alpha Detector} (EPEAD) and the \textit{High Energy Proton and Alpha Detector} (HEPAD) onboard GOES-13 and -15 (e.g. \citealp{ref:ONSAGER1996}). In case of event-integrated intensities, we also consider the high-energy (from 80\,MeV up to a few GeV) data collected by the PAMELA magnetic spectrometer \citep{ref:BRUNO2018}. 

The GOES particle detectors are known to be affected by a large background that makes them insensitive to relatively small SEP events, such as many of the events listed by \citet{ref:RICHARDSON2014}. On the other hand, they do not suffer from issues related to data gaps and signal saturation during large events. To improve the reliability of their spectroscopic observations we take advantage of the \textit{calibrated} energies derived by \citet{ref:SANDBERG2014} for the P2\,--\,P5 ($<$80\,MeV) proton energy channels, and by \citet{ref:BRUNO_GOES} for the P6\,--\,P11 ($>$80\,MeV) channels, based on the comparison with the STEREO and the PAMELA data, respectively.

\begin{sidewaysfigure}\center
\includegraphics[width=1\linewidth]{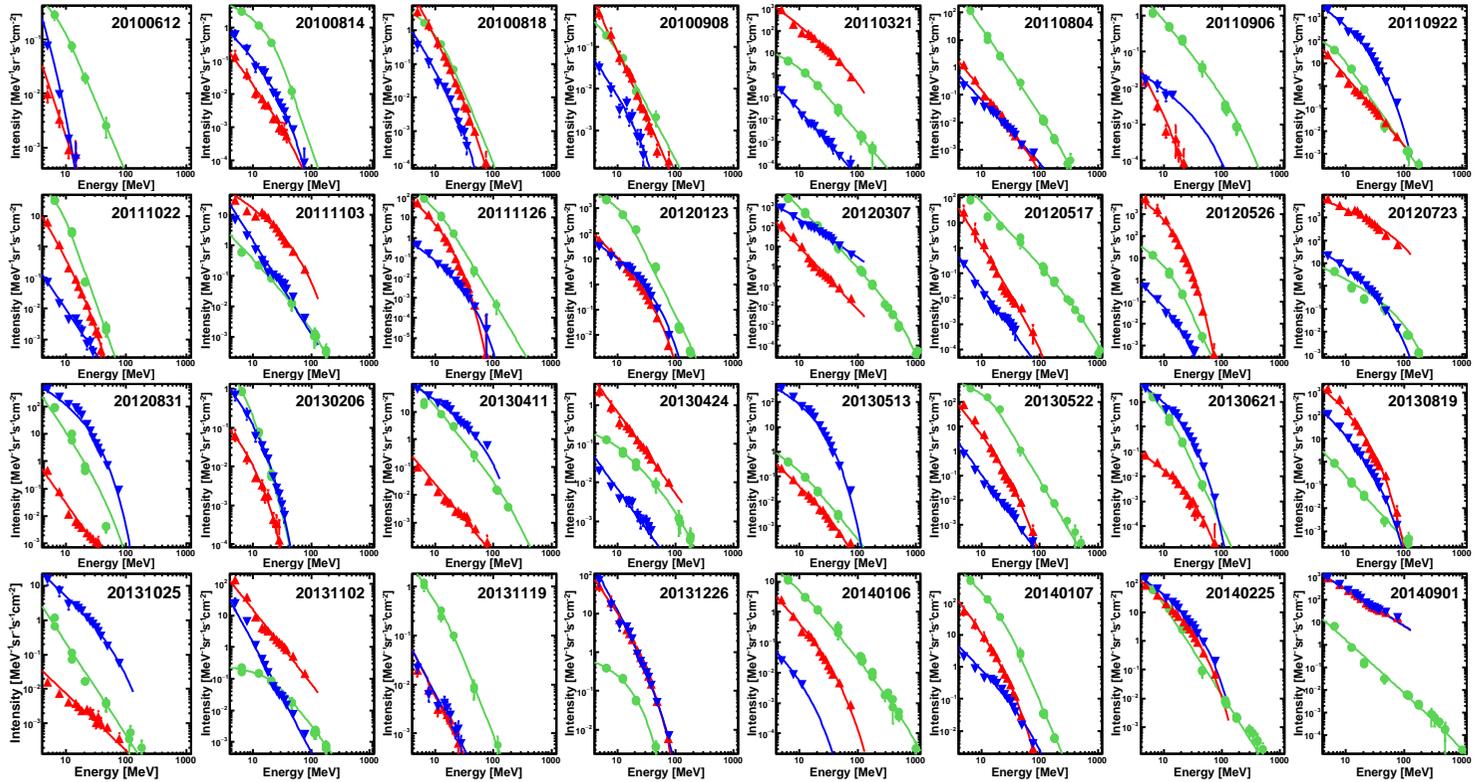}
\caption{Peak intensity spectral fits above 4\,MeV based on measurements by GOES-13/15 (green circles), STEREO-A (upward-pointing red triangles) and STEREO-B (downward-pointing blue triangles) during the 32 SEP events reported in Table \ref{tab:EventList}. Each panel reports the date of the corresponding SEP event.}
\label{fig:peak_spectra}
\end{sidewaysfigure}

\begin{sidewaysfigure}\center
\includegraphics[width=1\linewidth]{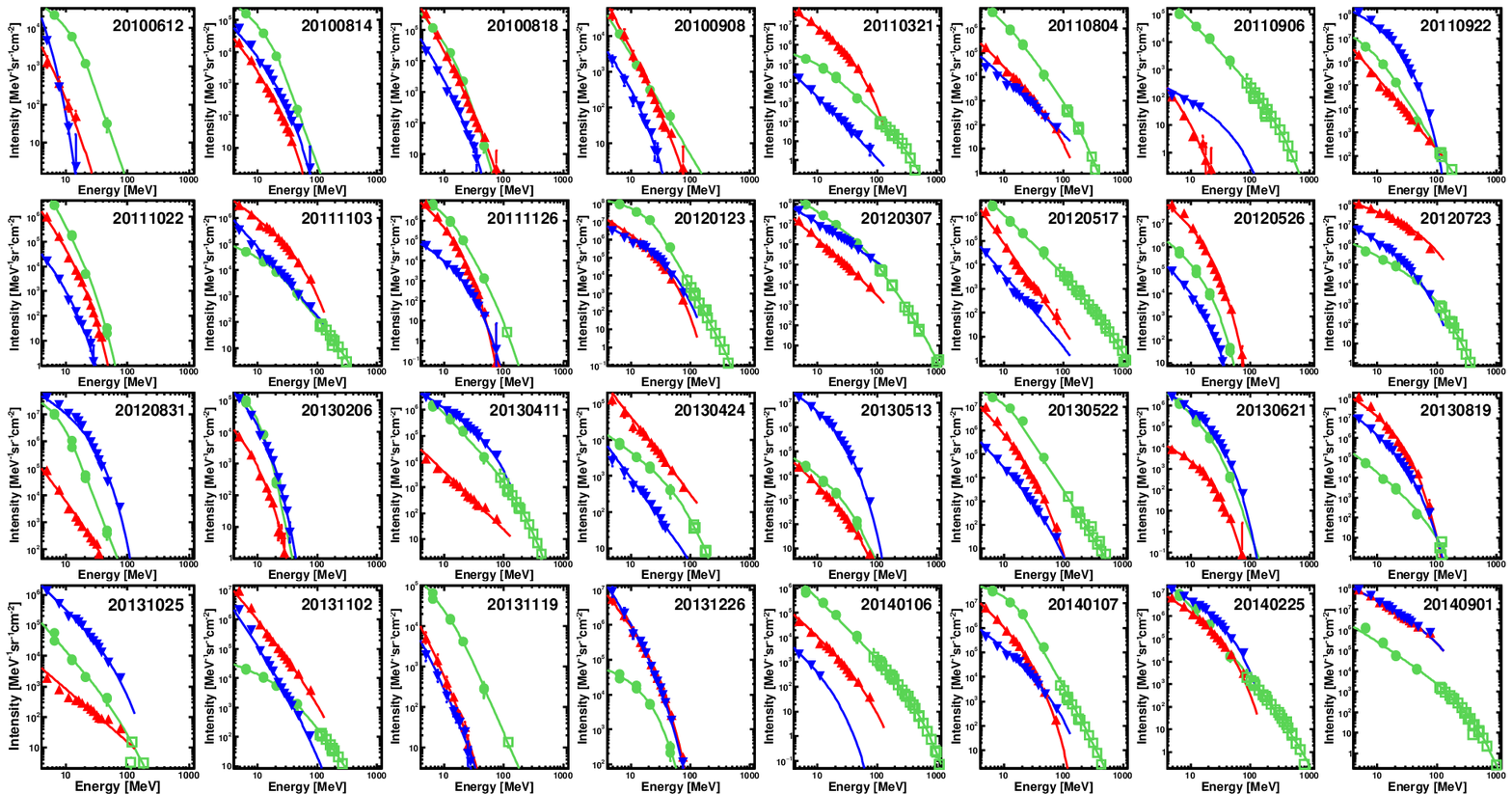}
\caption{Event-integrated intensity spectral fits above 4\,MeV based on measurements by GOES-13/15 (green circles), and PAMELA ($>$80\,MeV, green squares), STEREO-A (upward-pointing red triangles) and STEREO-B (downward-pointing blue triangles) during the 32 SEP events reported in Table \ref{tab:EventList}. Each panel reports the date of the corresponding SEP event.} 
\label{fig:fluence_spectra}
\end{sidewaysfigure}

SEP intensities are corrected for the quiet-time background (including the galactic cosmic-ray component), estimated from the minimum hourly running averages of intensities registered during a two-month interval prior to the event onset. STEREO spectra are fitted with the \citet{ref:ER1985} functional form, given by a power law with an exponential cutoff; near-Earth spectra, extending up to $\approx$1\,GeV, are fitted with a double power law with an exponential cutoff (see \citealp{ref:BRUNO2019} for details), that accounts for the high-energy spectral rollover reported by the PAMELA mission at several tens or hundreds of MeV \citep{ref:BRUNO2018}. The SEP peak and event-integrated intensity spectra above 4\,MeV reconstructed at the three spacecraft locations are reported in Figures \ref{fig:peak_spectra} and \ref{fig:fluence_spectra}, respectively. The spectral fits used in this work are limited to 130\,MeV, including a modest extrapolation from the STEREO observations (nominally limited to 100\,MeV), to minimize the significant uncertainties affecting the extension to higher energies, especially for hard spectra. In the case of the event \#32, large gaps present in STEREO-A data preclude a reliable spectrum reconstruction; to a first approximation, based on the comparison of the respective temporal profiles and the relatively small ($\approx$32$^\circ$) longitudinal separation of the two spacecraft, the SEP intensities at STEREO-A are assumed to be equal to those measured by STEREO-B.

Note that measured intensities may include contributions of particles that are locally accelerated at interplanetary shocks -- the so-called energetic storm particle (ESP) events. In particular, the resulting peak spectra for a given SEP event may include a mix of ESP-associated and ``first peak'' maxima, depending on energy and location. Also, particles from a previous event may be present at a particular spacecraft, or the onset of a new event may constrain the integration interval. These effects may influence the event-integrated intensities, especially for lower energies for which the event durations are longer, but the peak intensities should not be significantly affected since, for the event to be detected, it must significantly exceed the intensity of an ongoing event. 
Another possible source of uncertainty is related to Forbush-decrease effects, which reduce the particle intensity over a large energy range (e.g. \citealp{ref:SANDERSON1990}).

\section{SEP Intensity Spatial Distribution}\label{SEP intensity spatial distribution}
We assume that the peak or event-integrated intensity as a function of proton energy and location can be expressed as
\begin{equation}\label{phi_eq}
\Phi(E,\delta) = \Phi_{\mathrm{o}}(E)\hspace{1mm}G(E,\delta)
\end{equation}
where $\Phi_{\mathrm{o}}$ is the maximum intensity and $G$ is the normalized 2D spatial distribution at 1\,AU in terms of the great-circle or spherical distance from the location of the SEP 
distribution peak ($\alpha_{\mathrm{sep}}$, $\beta_{\mathrm{sep}}$):
\begin{equation}\label{eq:great-circle}
\delta = \arccos\left[ \sin(\alpha)\sin(\alpha_{\mathrm{sep}}) + \cos(\alpha)\cos(\alpha_{\mathrm{sep}})\cos(\beta-\beta_{\mathrm{sep}})\right],
\end{equation}
with $\alpha$ and $\beta$ being the HGS latitude and longitude.  The SEP intensities measured at the three locations -- which, for simplicity, are assumed to be isotropic -- are, following \citet{ref:COHEN2017} and \citet{ref:DENOLFO2019}, fitted by a periodic Gaussian function with form:
\begin{equation}\label{eq:periodic_Gaussian}
G(E,\delta) = \frac{1}{3}\left\{\exp\left[-\frac{\delta^{2}(E)}{2\sigma_{\mathrm{sep}}^{2}(E)}\right] + \exp\left[-\frac{\delta_{+}^{2}(E)}{2\sigma_{\mathrm{sep}}^{2}(E)}\right] + \exp\left[-\frac{\delta_{-}^{2}(E)}{2\sigma_{\mathrm{sep}}^{2}(E)}\right]\right\},
\end{equation}
where $\sigma_{\mathrm{sep}}$ is the distribution standard deviation assuming a spherical symmetry ($\sigma_{\mathrm{sep}}^{\alpha}$=$\sigma_{\mathrm{sep}}^{\beta}$=$\sigma_{\mathrm{sep}}$); the terms associated with
\begin{eqnarray}\label{eq:great-circle-pm}
\delta_{\pm}(E) &=& \arccos\left[ \sin(\alpha)\sin(\alpha_{\mathrm{sep}}(E)) + \right.\\ \nonumber
&&\left. \cos(\alpha)\cos(\alpha_{\mathrm{sep}}(E))\cos(\beta-\beta_{\mathrm{sep}}(E)\pm2\pi)\right]
\end{eqnarray}
ensure that 
$G(\delta)$ = $G(\delta\pm2\pi)$, 
accounting for the contribution from particles propagating at angles $>$180$^\circ$ from the center of the distribution, which is expected to be non-negligible for widespread events -- the use of a non-periodic Gaussian function would lead to an overestimate of the distribution width.
The parameters $\alpha_{\mathrm{sep}}$, $\beta_{\mathrm{sep}}$, and $\sigma_{\mathrm{sep}}$ are functions of particle energy. To a first approximation, $\alpha_{\mathrm{sep}}$ is assumed to coincide with the latitude of the parent CME direction ($\alpha_{\mathrm{sep}}$=$\alpha_{\mathrm{cme}}$) or with the flare latitude ($\alpha_{\mathrm{sep}}$=$\alpha_{\mathrm{flare}}$), if the CME information is not available.

Particle transport along the IMF is accounted for by computing the HGS location ($\alpha=\alpha_{\mathrm{sc}}$, $\beta=\beta_{\mathrm{sc}}$) of the footpoints of the Parker spiral field lines linking each spacecraft (hereafter referred as ``spacecraft footpoints''), which are mapped ballistically back to 2.5\,R$_{\odot}$ (see Appendix \ref{Appendix A}). This value roughly corresponds to the nominal source-surface height beyond which the spiral-model approximation is reasonable (see \citealp{ref:LEE2011} and references therein) and, moreover, to the radial distance where highest-energy particles are released from the CME-driven shock \citep{ref:REAMES2009,ref:GOPALSWAMY2013}. Thus it is unnecessary to consider using a coronal field model to map these field lines down to their photospheric footpoints. 
The footpoint calculation is performed at the CME first appearance time, by using the average solar-wind speed measured in the previous 30-minute interval.

In general, as a result of the Sun's rotation axis' tilt of about 7.25$^\circ$ from perpendicular to the Ecliptic plane, the spacecraft-footpoint locations do not lie in the solar equatorial plane ($\alpha_{\mathrm{sc}}$$\ne$0). Accordingly, projection effects are accounted for by multiplying measured SEP intensities by
\begin{equation}\label{footlat_corr}
K_{\alpha}(E) = \exp\left[ \frac{ \delta_{\mathrm{sc}}^{2}(E)-\delta_{\mathrm{sc,0}}^{2}(E)}{2\sigma_{\mathrm{sep}}^{2}(E)} \right],
\end{equation}
where $\delta_{\mathrm{sc}}$ and $\delta_{\mathrm{sc,0}}$ are the great-circle distances of the peak of the SEP spatial distribution from the spacecraft-footpoint location ($\alpha$=$\alpha_{\mathrm{sc}}$, $\beta$=$\beta_{\mathrm{sc}}$) and from their projection in the solar equatorial plane ($\alpha$=0, $\beta$=$\beta_{\mathrm{sc}}$), respectively.
We note that, even though the footpoint latitude is typically small, the differences in terms of great-circle distances can be much larger and the associated correction cannot be neglected. 

Finally, the spatial distribution parameters $\Phi_{\mathrm{o}}(E)$, $\beta_{\mathrm{sep}}(E)$ and $\sigma_{\mathrm{sep}}(E)$ are derived from the corrected spacecraft measurements by using the projection of Equation \ref{eq:periodic_Gaussian} in the solar equatorial plane:
\begin{equation}\label{eq:ProjDistr}
\Phi_{\mathrm{eq}}(E; \beta) = \Phi_{\mathrm{o}}(E) \hspace{0.5mm} G_{\mathrm{eq}}(E; \beta), \end{equation}
where 
\begin{equation}\label{eq:GaussianProj}
G_{\mathrm{eq}}(E; \beta) = \frac{1}{3} \hspace{0.5mm} \exp\left\{ -\frac{\arccos^{2}\left[\cos(\alpha_{\mathrm{cme}})\cos(\beta-\beta_{\mathrm{sep}}(E))\right]}{2\sigma_{\mathrm{sep}}^{2}(E)}\right\} + . . .,
\end{equation}
with the two terms centered at $\beta_{\mathrm{sep}}\pm2\pi$ omitted here for brevity. 
Since $\sigma_{\mathrm{sep}}$ is unknown a priori, the correction factor given by Equation \ref{footlat_corr} is obtained through an iterative procedure, using $\sigma_{\mathrm{sep}}$=43$^\circ$ as initial value \citep{ref:RICHARDSON2014}, until the fit results become stable.

\begin{sidewaysfigure}\center
\includegraphics[width=1\linewidth]{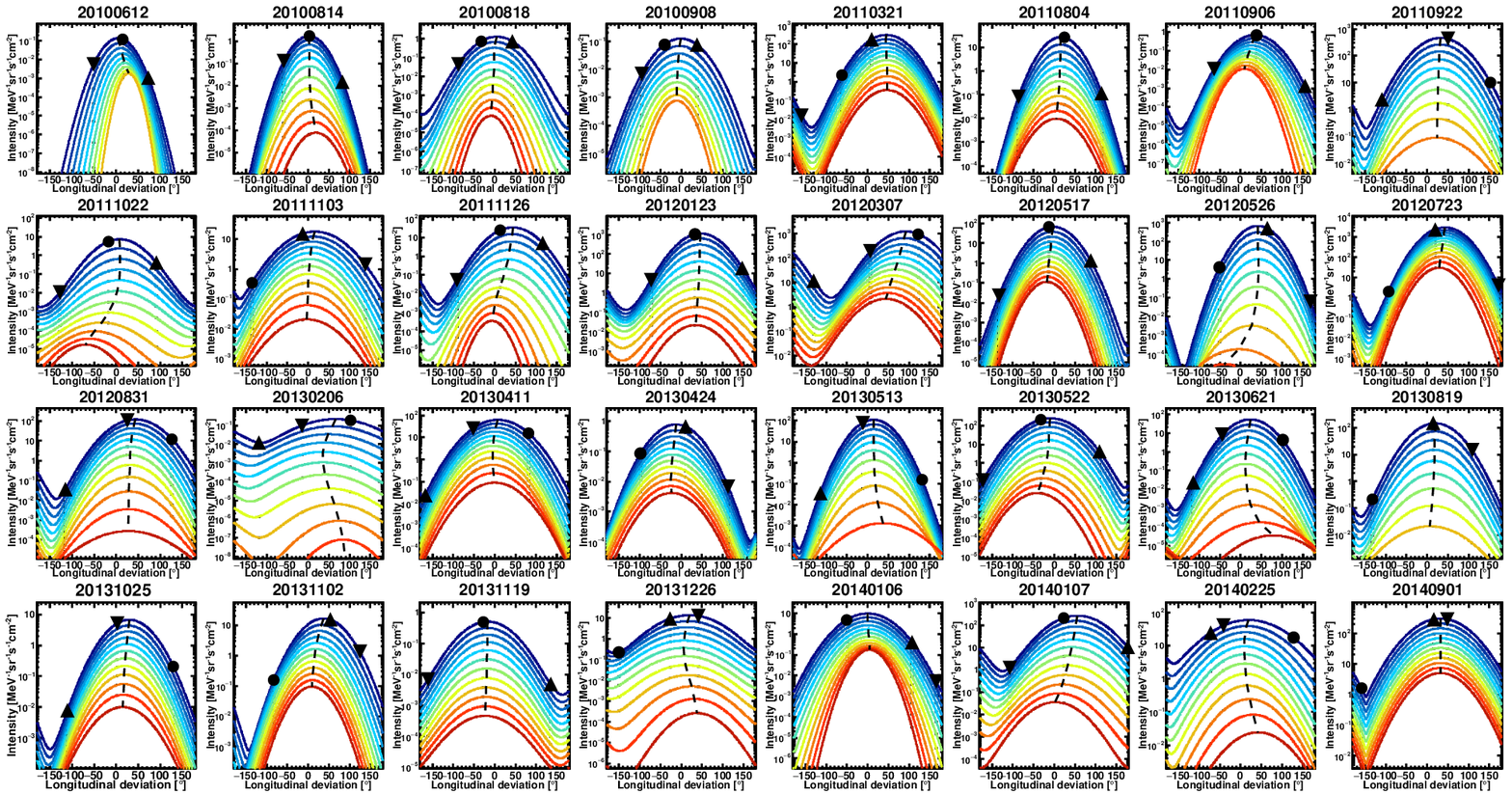}
\caption{1\,AU spatial distribution of the 32 SEP events used in the analysis (Equation \ref{eq:ProjDistr}) determined from the fits of the peak intensities measured by GOES (circles) and STEREO-A/B (triangles), as a function of the longitudinal deviation between the spacecraft magnetic footpoints (estimated at CME onset and at 2.5\,R$_{\odot}$ height) and the CME direction. Each panel corresponds to a different SEP event in Table \ref{tab:EventList}. The color code refers to 12 logarithmically spaced energy values between 10 and 130\,MeV. The distributions are centered at the peak connection angle evaluated at 10\,MeV; the dashed lines are the curves linking the distribution peaks at different energies.}
\label{fig:peak_fit}
\end{sidewaysfigure}

\begin{sidewaysfigure}\center
\includegraphics[width=1\linewidth]{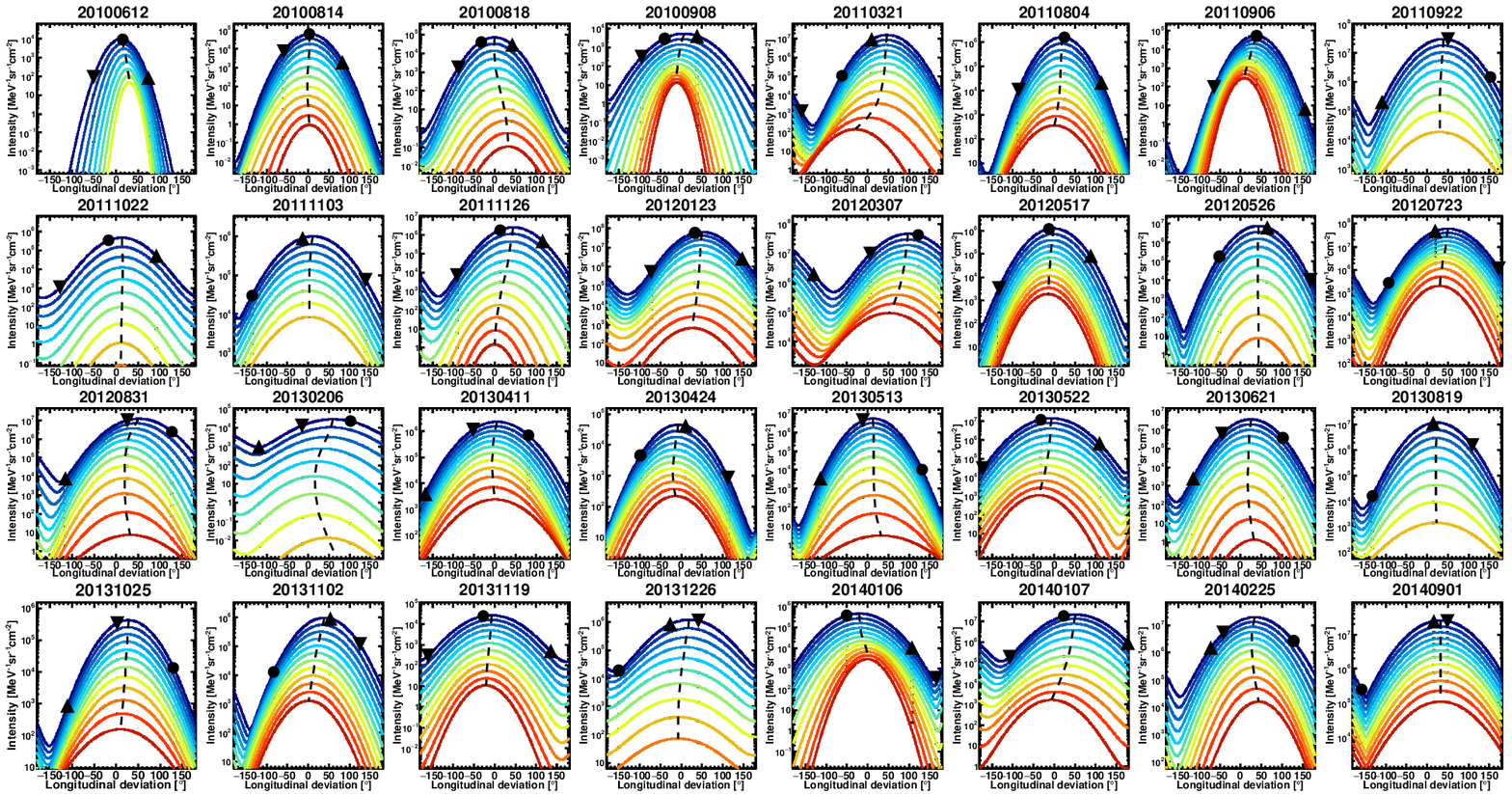}
\caption{1\,AU spatial distribution of the 32 SEP events used in the analysis (Equation \ref{eq:ProjDistr}) determined from the fits of the event-integrated intensities measured by GOES/PAMELA (circles) and STEREO-A/B (triangles), as a function of the longitudinal deviation between the spacecraft magnetic footpoints (estimated at CME onset and at 2.5\,R$_{\odot}$ height) and the CME direction. Each panel corresponds to a different SEP event in Table \ref{tab:EventList}. The color code refers to 12 logarithmically spaced energy values between 10 and 130\,MeV. The distributions are centered at the peak connection angle evaluated at 10\,MeV; the dashed lines are the curves linking the distribution peaks at different energies.} 
\label{fig:fluence_fit}
\end{sidewaysfigure}

\subsection{Results}
Results relative to peak and event-integrated intensity distributions as a function of the longitudinal connection angle [$\beta_{\mathrm{sep}}(E) - \beta_{\mathrm{cme}}$] for the 32 analyzed SEP events are shown in Figures \ref{fig:peak_fit} and \ref{fig:fluence_fit}, respectively. Negative/positive angles correspond to locations eastward/westward of the CME direction. Depending on the associated uncertainties, SEP spectral fits shown in Figures \ref{fig:peak_spectra} and \ref{fig:fluence_spectra} are interpolated at up to 12 logarithmically spaced energy values between 10 and 130\,MeV (indicated by the line color) and used to reconstruct the corresponding spatial distributions. The three markers denote the longitudinal deviation of the spacecraft footpoints (circles=GOES/PAMELA, triangles=STEREO-A/B). The dashed lines are the curves linking the distribution peaks at different energies. Both of the locations of the peaks and the widths of the distributions show some energy dependences that vary from event to event and deserve further investigation in a separate study.

 \begin{figure}[!t]\center
\includegraphics[width=\linewidth]{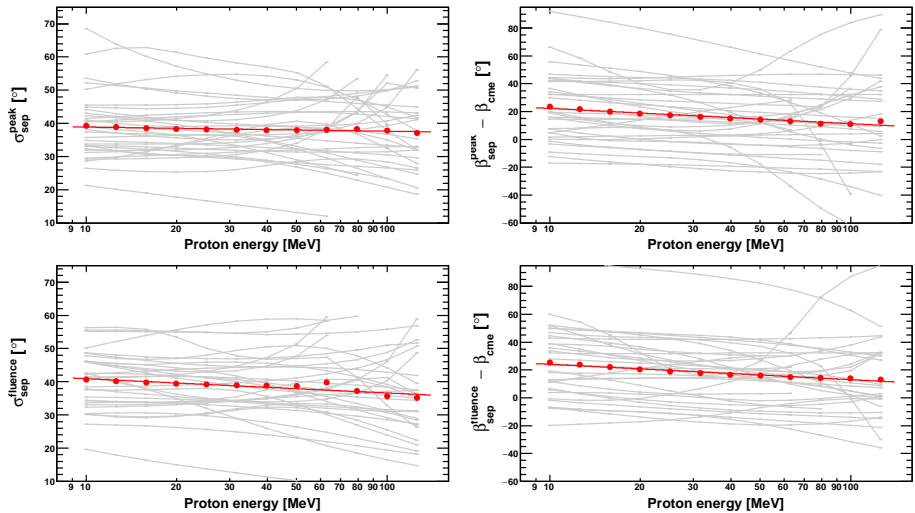}
\caption{Energy dependence of the spatial distribution standard deviation [$\sigma_{\mathrm{sep}}$; left panels] and of the longitudinal deviation [$\beta_{\mathrm{sep}}$-$\beta_{\mathrm{cme}}$; right panels] from the Gaussian fits of peak and event-integrated intensities (top and bottom panels, respectively). Each gray line corresponds to a different SEP event. The red points are the values averaged over the whole SEP event sample, with the red lines the corresponding fits based on Equations \ref{eq:beta}\,--\,\ref{eq:sigma}.} 
\label{fig:sigmabeta}
\end{figure}

Figure \ref{fig:sigmabeta} shows the energy distributions of the related $\beta_{\mathrm{sep}}$ and $\sigma_{\mathrm{sep}}$ parameters, for both peak (top panels) and event-integrated (bottom panels) intensities. Significant event-to-event variations can be noted. 
Investigating their causes is beyond the scope of this study but they might, for example, include contributions from overlapping events or local shock-associated particles at certain spacecraft and energies. In addition, the connection angle of a given spacecraft might be influenced by the presence of interplanetary magnetic structures between the source and the spacecraft location. While such effects introduce significant uncertainties in the estimates of the spatial extent of some SEP events, they are an integral part of the average SEP event properties observed at 1\,AU and, as such, are included when constructing the empirical model.

For each proton energy, mean values $\bar{\beta}_{\mathrm{sep}}$ and $\bar{\sigma}_{\mathrm{sep}}$ are obtained by averaging over the selected SEP event sample (red points). In order to parameterize their energy dependence, the mean-value distributions are fitted with the functions
\begin{equation}\label{eq:beta}
\bar{\beta}_{\mathrm{sep}}(E) - \beta_{\mathrm{cme}} = \beta_{0} - \beta_{1}\log(E),
\end{equation}
where $\beta_{\mathrm{cme}}$ is the longitude associated with the CME direction
-- or the flare longitude if the CME information is not available -- and
\begin{equation}\label{eq:sigma}
\bar{\sigma}_{\mathrm{sep}}(E) = \sigma_{0} - \sigma_{1}\log(E).
\end{equation}
The derived best-fits are denoted by the red curves in Figure \ref{fig:sigmabeta}, with the corresponding parameters ($\beta_{0}$, $\beta_{1}$) and ($\sigma_{0}$, $\sigma_{1}$) reported in Table \ref{spatialpar_table}. 

\begin{table}[!t]
\center
\begin{tabular}{lcccc}
 & $\beta_{0}$ & $\beta_{1}$ & $\sigma_{0}$ & $\sigma_{1}$ \\
\hline
Peak intensities & 3.38$\times$10$^{1}$ & 5.11$\times$10$^{0}$ & 4.05$\times$10$^{1}$ & 5.20$\times$10$^{-1}$ \\ 
Event-int. intensities & 3.52$\times$10$^{1}$ & 4.81$\times$10$^{0}$ & 4.52$\times$10$^{1}$ & 1.87$\times$10$^{0}$ \\ 
\hline
\end{tabular}
\caption{Best-fit parameters for Equations \ref{eq:beta}\,--\,\ref{eq:sigma} derived for both peak and event-integrated intensities.}
\label{spatialpar_table}
\end{table}

Consistent with previous studies (e.g. \citealp{ref:COHEN2017}), the standard deviation $\bar{\sigma}_{\mathrm{sep}}$ is found to decrease with increasing proton energy, although the variation is relatively small for peak intensities. In addition, the peak of the spatial distribution is, on average, located on field lines with footpoints located westward of the CME direction, and tends to move eastward with increasing energy, so that highest-energy protons are found on field lines with footpoints close to the CME direction. 

%
%
%
%
%
%
%
%
%
%
%
\section{Dependence of SEP Intensity on CME Speed}\label{Dependence on CME speed}
Similar to \citet{ref:RICHARDSON2014} and \citet{ref:RICHARDSON2018}, we assume that SEP intensity is correlated with the parent CME speed.
We do not consider the CME angular widths as they tend to be correlated with the speeds (e.g. Figure 8 of \citealp{ref:RICHARDSON2015}).
Since the SEP intensity--CME speed correlation is energy-dependent (as shown below), we can include this dependence in the parameterization of the SEP spatial distribution maximum intensity $\Phi_{\mathrm{o}}(E)=\Phi_{\mathrm{o}}(E,V_{\mathrm{cme}})$,
according to the formula
\begin{equation}\label{eq:cme1}
\Phi_{\mathrm{o}}(E,V_{\mathrm{cme}}) = \Psi_{\mathrm{cme}}(E) \hspace{1mm} \exp\left(\Lambda_{\mathrm{sep}}(E) \hspace{1mm} V_{\mathrm{cme}}\right),
\end{equation}
where 
\begin{equation}\label{eq:cme2}
\Psi_{\mathrm{cme}}(E) = \psi_{0} \left(E/E_{0}\right)^{-\psi_{1}}\exp\left(-E/E_{r}\right)
\end{equation}
and 
\begin{equation}\label{eq:cme3}
\Lambda_{\mathrm{cme}}(E) = \lambda_{0} \left(E/E_{0}\right)^{\lambda_{1}}
\end{equation}
account for the energy dependence, 
with $E_{0}$=10\,MeV the threshold energy and $E_{r}$=300\,MeV the cutoff energy.
The spectral form in Equation \ref{eq:cme2} is that proposed by \citet{ref:ER1985};
the $E_{r}$ value is assumed ad hoc to reproduce the SEP spectral rollover reported at high energies by the PAMELA experiment (see \citealp{ref:BRUNO2018} for details), providing a reasonable description of intensities beyond 130\,MeV.

\begin{figure}[!t]
\center
\includegraphics[width=\textwidth]{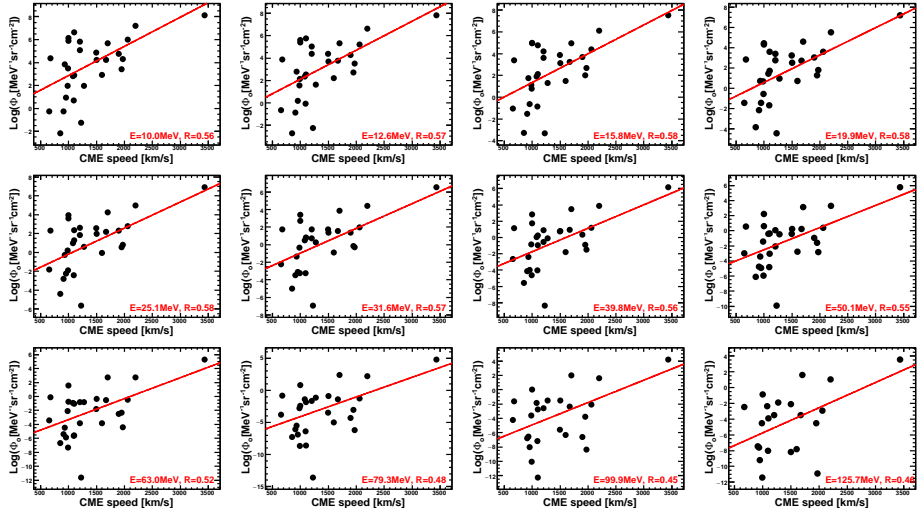} 
\caption{Distribution of SEP peak-intensity maxima [$\log(\Phi_{\mathrm{o}})$] as a function of CME speed [$V_{\mathrm{cme}}$], for 12 energy values between 10 and 130\,MeV. The corresponding regression lines are also reported, along with the correlation factor $R$ values.} 
\label{fig:peak4}
\end{figure}
\begin{figure}[!h]
\center
\includegraphics[width=\textwidth]{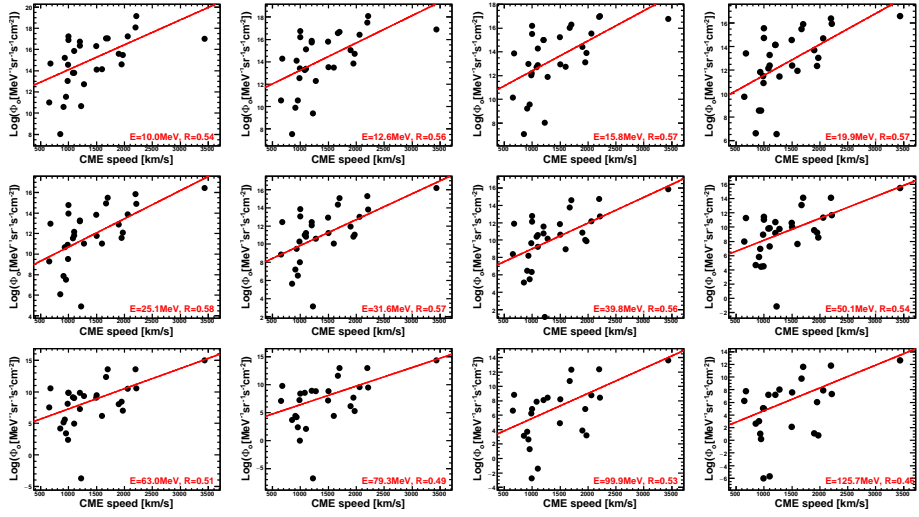} 
\caption{Distribution of SEP event-integrated-intensity maxima [$\log(\Phi_{\mathrm{o}})$] as a function of CME speed [$V_{\mathrm{cme}}$], for 12 energy values between 10 and 130\,MeV. The corresponding regression lines are also reported, along with the correlation factor $R$ values.} 
\label{fig:fluence4}
\end{figure}

\begin{figure}[!t]
\center
\includegraphics[width=\linewidth]{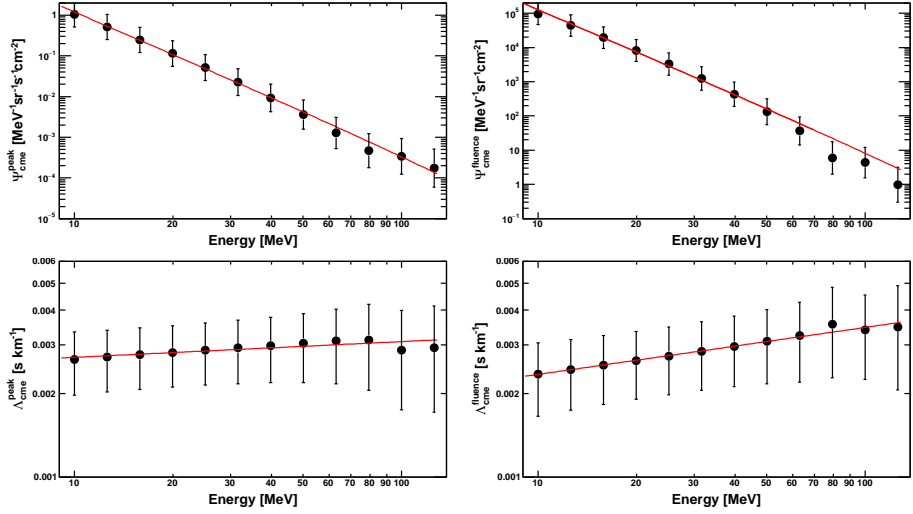}
\caption{Distribution of $\Psi_{\mathrm{cme}}$ and $\Lambda_{\mathrm{cme}}$ values (top and bottom panels, respectively) from the individual fits performed at different proton energies, for peak intensities (left panels, from Figure \ref{fig:peak4}) and event-integrated intensities (right panels, from Figure \ref{fig:fluence4}). The red curves represent the fits with Equations \ref{eq:cme2}-\ref{eq:cme3}.}
\label{fig:Acme}
\end{figure}

The free parameters ($\psi_{0}$,$\psi_{1}$) and ($\lambda_{0}$,$\lambda_{1}$) are estimated as follows. The calculation involves 29 out of the 32 selected SEP events reported in Table \ref{tab:EventList} for which the CME information is available.
First, for each energy value $i$=1,12, 
a linear fit of
the $\log(\Phi_{\mathrm{o,i}})$ versus $V_{\mathrm{cme}}$ distribution is 
performed, deriving the individual $\Psi_{\mathrm{cme},i}$ and $\Lambda_{\mathrm{cme},i}$ parameters (see Figures \ref{fig:peak4} and \ref{fig:fluence4}). 
As often noted in other studies of SEP intensities versus CME speeds, there is a large spread about the best-fit line at each energy for this sample of events, but the general trends are evident. At some energies, the fit appears to be dominated by event \#16 associated with an unusually fast CME, but 
it is also consistent with the trend evident in the remaining events.
Then, the distributions of the $\Psi_{\mathrm{cme},i}$ and $\Lambda_{\mathrm{cme},i}$ values as a function of proton energy are fitted by using the functional forms in Equations \ref{eq:cme2} and \ref{eq:cme3} respectively, as shown in Figure \ref{fig:Acme}, providing an estimate of the best-fit parameters ($\psi_{0}$,$\psi_{1}$) and ($\lambda_{0}$,$\lambda_{1}$), reported in Table \ref{tab:cme}.

\begin{table}[!t]
\center
\begin{tabular}{lcccc}
 & $\psi_{0}$ & $\psi_{1}$ & $\lambda_{0}$ & $\lambda_{1}$ \\
\hline
Peak intensities & 1.50$\times$10$^{0}$ & 3.61$\times$10$^{0}$ & 2.55$\times$10$^{-3}$ & 9.01$\times$10$^{-2}$ \\ 
Event-int. intensities & 1.31$\times$10$^{5}$ & 4.06$\times$10$^{0}$ & 2.35$\times$10$^{-3}$ & 1.69$\times$10$^{-1}$ \\ 
\hline
\end{tabular}
\caption{Best-fit parameters for Equations \ref{eq:cme2} and \ref{eq:cme3} derived for both peak and event-integrated intensities.}
\label{tab:cme}
\end{table}

\section{Model Uncertainties}\label{Model uncertainties}
The uncertainty in the predicted SEP event-integrated or peak intensity $\Phi(E,V_{\mathrm{cme}})$ can be obtained by propagating the errors on
the spacecraft footpoint location, 
the CME speed, and the energy-dependent $\bar{\sigma}_{\mathrm{sep}}$, $\bar{\beta}_{\mathrm{sep}}$, $\Psi_{\mathrm{cme}}$ and $\Lambda_{\mathrm{cme}}$ values given by Equations \ref{eq:beta}\,--\,\ref{eq:sigma} and \ref{eq:cme2}\,--\,\ref{eq:cme3}, according to the standard formula:
\begin{eqnarray}
\delta\Phi &=&\biggl\{
\left[\frac{\partial \Phi}{\partial \bar{\sigma}_{\mathrm{sep}}}\delta\bar{\sigma}_{\mathrm{sep}}\right]^{2} + 
\left[\frac{\partial \Phi}{\partial \alpha_{\mathrm{cme}}}\delta\alpha_{\mathrm{cme}}\right]^{2} +
\left[\frac{\partial \Phi}{\partial \bar{\beta}_{\mathrm{sep}}}\delta\bar{\beta}_{\mathrm{sep}}\right]^{2} + \\ \nonumber
&& \left[\frac{\partial \Phi}{\partial \alpha_{\mathrm{sc}}}\delta\alpha_{\mathrm{sc}}\right]^{2} +
\left[\frac{\partial \Phi}{\partial \beta_{\mathrm{sc}}}\delta\beta_{\mathrm{sc}}\right]^{2} + \\ \nonumber
&&\left[\frac{\partial \Phi}{\partial \Psi_{\mathrm{cme}}}\delta\Psi_{\mathrm{cme}}\right]^{2} + 
\left[\frac{\partial \Phi}{\partial \Lambda_{\mathrm{cme}}}\delta\Lambda_{\mathrm{cme}}\right]^{2} + 
\left[\frac{\partial \Phi}{\partial V_{\mathrm{cme}}}\delta V_{\mathrm{cme}}\right]^{2}
\biggl\}^{\!1/2}.
\end{eqnarray}
All parameters associated with the aforementioned quantities 
(e.g. $\sigma_{0}$ and $\sigma_{1}$ in the case of $\bar{\sigma}_{\mathrm{sep}}$)
are assumed to be dependent, and related errors are estimated by using the corresponding covariance matrices. 
Conservatively, for CME parameter estimates based on three-spacecraft observations, a 5$^\circ$ and 10$^\circ$ error is assumed on $\alpha_{\mathrm{cme}}$ and $\beta_{\mathrm{cme}}$, respectively, while a 20\% uncertainty is associated with $V_{\mathrm{cme}}$; for two-spacecraft measurements (i.e. after the termination of STEREO-B), the uncertainties are 10$^\circ$, 15$^\circ$, and 30\%, respectively (L. Mays, private communication, 2020).
Finally, a 200\,MeV error is associated with $E_{r}$.
As discussed in Appendix \ref{Appendix A}, 
the uncertainty in the footpoint longitude accounts for a 100\,km s$^{-1}$ error associated with the solar-wind speed estimate and includes a 25$^\circ$ error accounting for interplanetary transport effects ignored by the model, such as 
the influence of
magnetic fluctuations (e.g. \citealp{ref:IPPOLITO2005}) 
and solar wind structures (e.g. \citealp{ref:LARIOKARELITZ2014,ref:MASSON2012}), or the propagation across the nominal Parker spiral magnetic field 
(e.g. \citealp{ref:JOKIPII1966,ref:DALLA2013,ref:LAITINEN2016}). Similarly, a 10$^\circ$ error is assumed on the footpoint latitude 
accounting for the deviations from the spiral model approximation out of the equatorial plane
and effects related to the differential changes across the solar surface. 
The total error $\delta\Phi$ increases with $V_{\mathrm{cme}}$, and it is dominated by the uncertainties in $\Psi_{\mathrm{cme}}$ and $\Lambda_{\mathrm{cme}}$, as a consequence of the spread in the $\Phi_{\mathrm{o}}$ versus $V_{\mathrm{cme}}$ distributions in Figures \ref{fig:peak4} and \ref{fig:fluence4}.

\section{Model Testing and Caveats}\label{Model testing}
In order to test the empirical SEP prediction model based on the parameterization of the sample of SEP events discussed in previous sections, we compare the observed and calculated spectra for 20 events not used to derive the model parameters. The selected sample, listed in Table \ref{tab:TestEventList}, essentially comprises energetic events detected between 2011 and 2017 at Earth and, possibly, at one STEREO spacecraft -- with the exception of the 30 September 2013 three-spacecraft event -- as a consequence of a relatively narrow spatial distribution compared to the spacecraft separation, a large background component from a previous event that precludes a direct measurement of the event, 
significant data gaps (e.g. between September 2014 and November 2015),
or after the loss of contact with STEREO-B in October 2014. This last subset includes the 10 September 2017 event, associated with the 72nd GLE \citep{ref:BRUNO2019}. 

\begin{table}[!t]
\center
\setlength{\tabcolsep}{4.9pt}
\begin{tabular}{ccrrcccc}
 & \multicolumn{4}{c}{CME} & \multicolumn{3}{c}{Spacecraft footpoints}\\
\# & Onset time & Speed & Width & Direction & STB & Earth & STA\\
\hline
1 & 2011-03-07T20:12 & 1980 & 90 & N17W50 & N01E16$^{b}$ & S07W55 & S00W153 \\ 
2* & 2011-06-07T06:50 & 1400 & 92 & S25W52 & S07E48$^{b}$ & N00W55 & N07W122$^{b}$ \\ 
3 & 2011-08-09T08:30 & 1175 & 40 & S12W62 & S03E10$^{a}$ & N06W42 & N02W138 \\ 
4* & 2012-01-27T16:39 & 2200 & 110 & N40W75 & N06E56$^{b}$ & S05W47 & S02W160 \\ 
5* & 2012-03-13T17:52 & 2250 & 120 & N25W43 & N02E63$^{b}$ & S07W37 & N03W162$^{b}$ \\ 
6* & 2012-07-06T23:12 & 1200 & 80 & S35W65 & S07E66$^{a}$ & N03W53 & N03W170 \\ 
7* & 2012-07-08T16:48 & 1000 & 60 & S17W74 & S07E58$^{a}$ & N03W56 & N03W177 \\ 
8* & 2012-07-12T16:54 & 1300 & 130 & S13W06 & S07E46 & N04W50 & N03W160$^{b}$ \\ 
9 & 2012-07-17T14:24 & 1100 & 90 & S30W54 & S06E73$^{a}$ & N04W55 & N02E174 \\ 
10* & 2012-07-19T05:36 & 1550 & 120 & S14W94 & S06E58$^{a}$ & N04W53 & N02E176$^{b}$ \\ 
11* & 2013-09-29T22:40 & 1100 & 140 & N26W38 & S02E46 & N06W87 & S07E129 \\ 
12* & 2013-10-28T14:12 & 1100 & 70 & N27W80 & N00E69 & N04W84$^{b}$ & S06E158$^{b}$ \\ 
13 & 2013-11-07T10:39 & 2100 & 184 & S13E135 & N00E101 & N04W67$^{b}$ & S06E171 \\ 
14 & 2013-12-28T18:00 & 769 & 130 & S01W101 & N05E73$^{b}$ & S02W79 & S00E151$^{b}$ \\ 
15 & 2014-02-20T08:00 & 854 & 80 & S04W79 & N07E81$^{a}$ & S07W45 & N05E164$^{b}$ \\ 
16* & 2014-04-18T13:09 & 1400 & 90 & S34W10 & N04E76$^{a}$ & S05W60 & N06E142$^{a}$ \\ 
17* & 2014-09-10T18:18 & 1400 & 90 & N15W10 & S06E95 & N07W66 & S07E129$^{b}$ \\ 
18 & 2015-06-21T02:48 & 1250 & 114 & N07E08 & ... & N01W74 & ... \\ 
19 & 2017-07-14T01:36 & 750 & 98 & S09W40 & ... & N04W58 & S07E86$^{a}$ \\ 
20 & 2017-09-10T16:09 & 2650 & 108 & S12W85 & ... & N07W46 & S04E69 \\ 
\hline
\end{tabular}
*Associated with SEP events observed by PAMELA \citep{ref:BRUNO2018}.
$^{a}$No significant SEP signal.
$^{b}$High background from an ongoing SEP event. 
\caption{List of CMEs associated with the SEP events used for testing the empirical model. The first column is the event number. Columns 2\,--\,5 report the CME first appearance time [UT], space speed [km s$^{-1}$], angular width [$^\circ$] and direction from the DONKI catalog. 
The next three columns list the location of the footpoints of the Parker spiral field lines crossed by each spacecraft (STEREO-A/B and GOES/PAMELA) at CME onset, mapped ballistically back to 2.5\,R$_{\odot}$. Footpoints locations and CME directions are expressed in terms of HGS latitudes/longitudes. The dots (...) indicate no STEREO data available.}
\label{tab:TestEventList}
\end{table}

\begin{figure}
\center
\includegraphics[width=\textwidth]{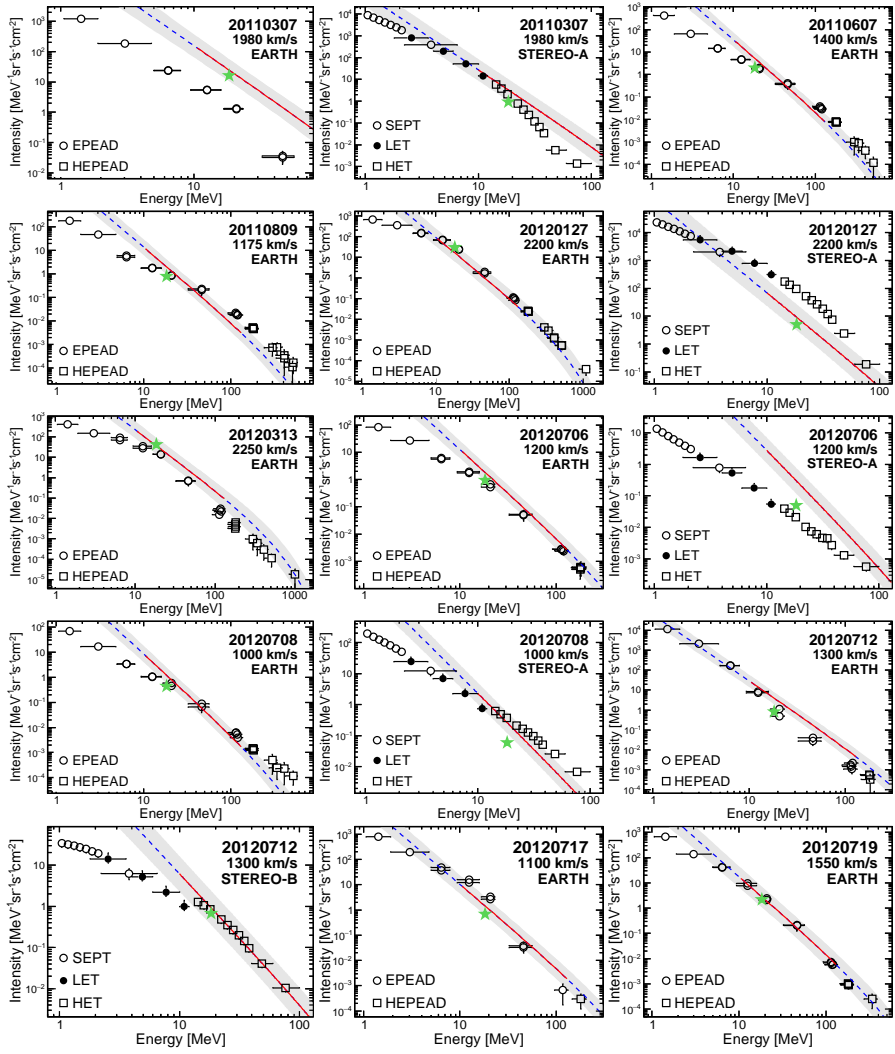}
\caption{Peak spectra of the test SEP events \#1\,--\,10 listed in Table \ref{tab:TestEventList},
measured at Earth and/or STEREO. 
The event date, the related CME speed and the spacecraft location are reported in each panel.
The red lines are the predicted spectra 
between 10\,--\,130\,MeV; the dashed blue lines are the extrapolations to lower/higher energies. 
The gray band denotes the one-$\sigma$ uncertainty associated with the model prediction.
For a comparison, the green stars are the 14\,--\,24\,MeV peak intensities based on the model by \citet{ref:RICHARDSON2018}.} 
\label{fig:testplot_peak1}
\end{figure}

\begin{figure}
\center
\includegraphics[width=\textwidth]{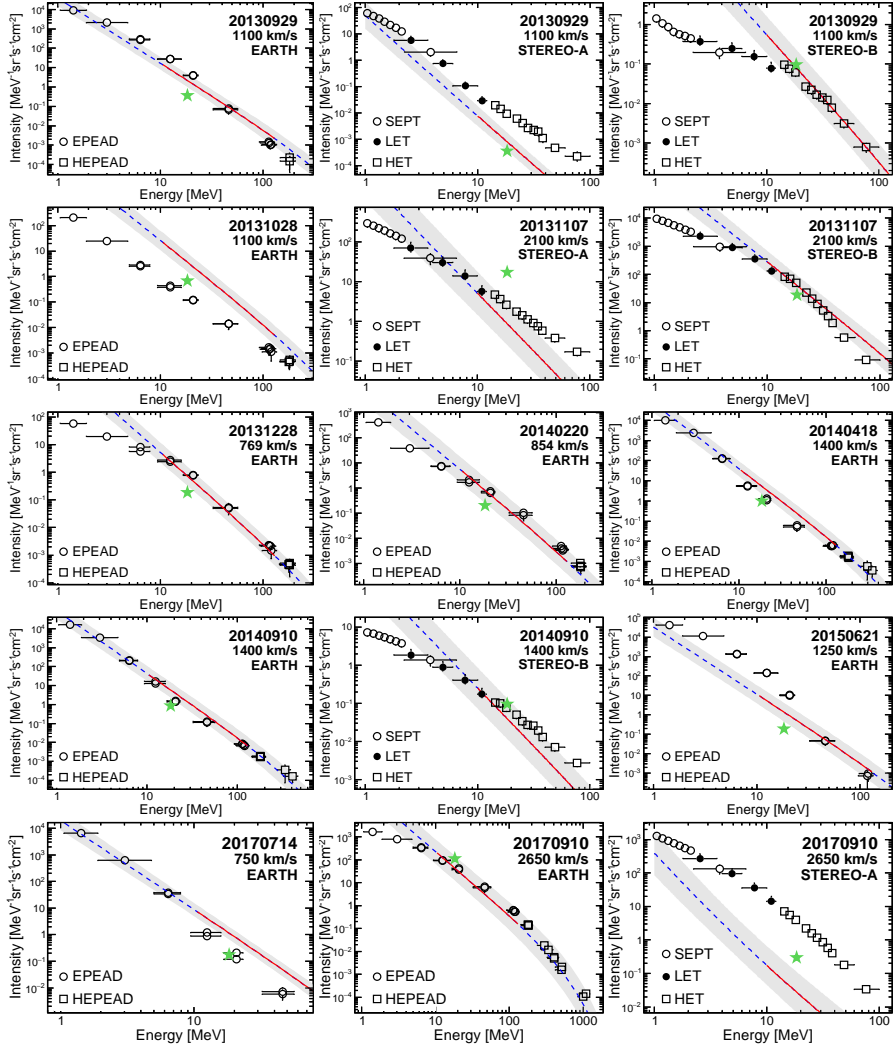}
\caption{Peak spectra of the test SEP events \#11\,--\,20 listed in Table \ref{tab:TestEventList}, in the same format as Figure \ref{fig:testplot_peak1}.} 
\label{fig:testplot_peak2}
\end{figure}

\begin{figure}
\center
\includegraphics[width=\textwidth]{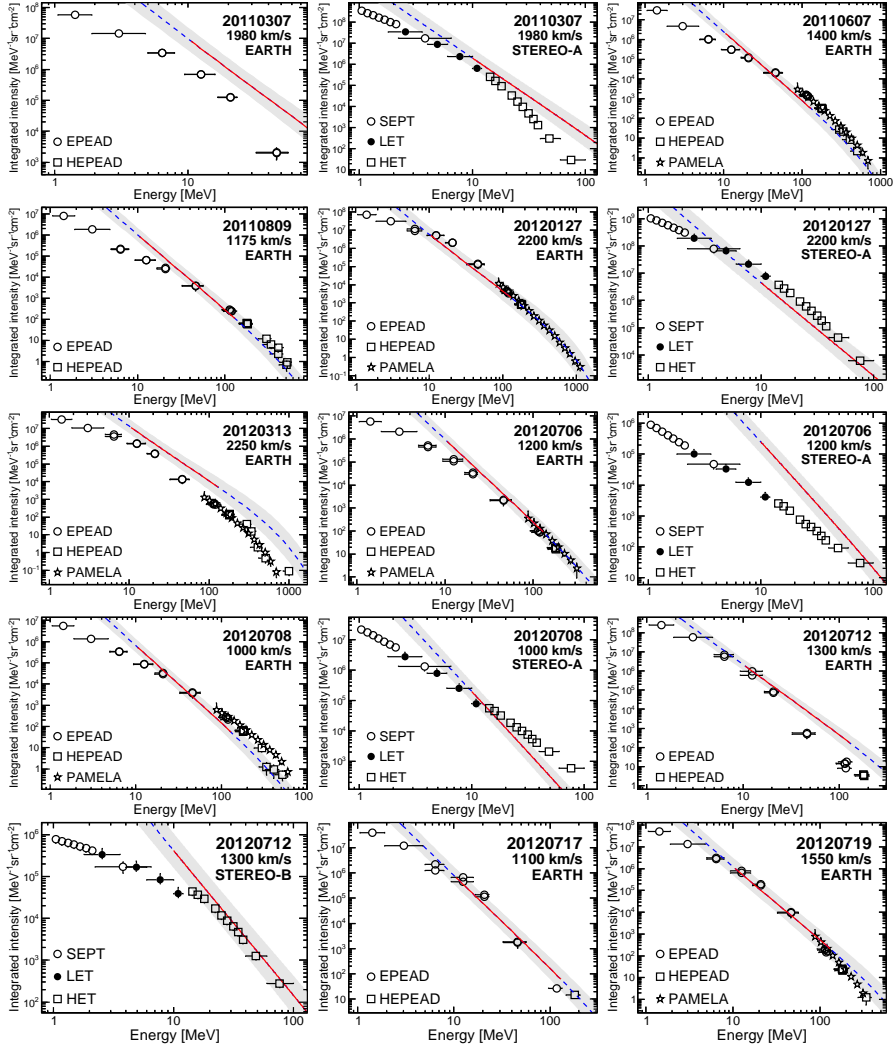}
\caption{Event-integrated spectra of the test SEP events \#1\,--\,10 listed in Table \ref{tab:TestEventList},
measured at Earth and/or STEREO (see Figure \ref{fig:testplot_peak1} for details).
In this case, the experimental data include the high-energy observations of the PAMELA experiment (when available), marked by empty stars.} 
\label{fig:testplot_fluence1}
\end{figure}

\begin{figure}
\center
\includegraphics[width=\textwidth]{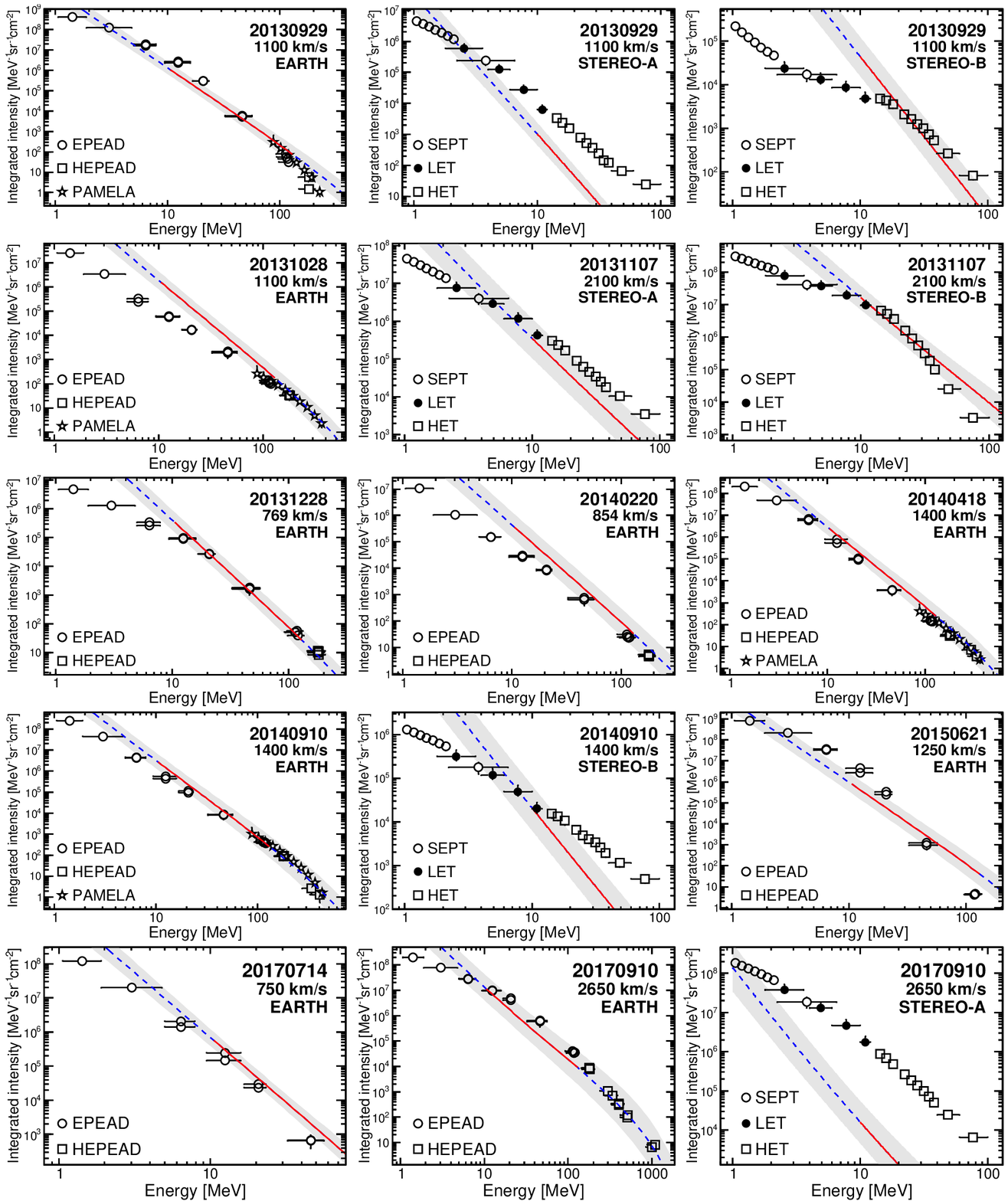}
\caption{Event-integrated spectra of the test SEP events \#11\,--\,20 listed in Table \ref{tab:TestEventList}, in the same format as Figure \ref{fig:testplot_fluence1}.} 
\label{fig:testplot_fluence2}
\end{figure}

The SEP peak and event-integrated spectra for these 20 events reconstructed at Earth and STEREO are displayed in Figures \ref{fig:testplot_peak1}\,--\,\ref{fig:testplot_peak2} and \ref{fig:testplot_fluence1}\,--\,\ref{fig:testplot_fluence2}, respectively. 
Near-Earth measurements, extending up to $\approx$1\,GeV, are based on the GOES-13/15 EPEAD/HEPAD sensors and, in case of event-integrated intensities, on the high-energy PAMELA data. STEREO observations, limited to 100\,MeV, rely on the SIT, LET, and HET detectors. The corresponding model predictions between 10\,--\,130\,MeV are marked by the solid red lines, while the extrapolation of each spectrum outside the nominal energy range is denoted by the dotted blue lines. The gray bands indicate the one-$\sigma$ uncertainty associated with the predicted spectra, 
as discussed above. For comparison, the green stars in Figures \ref{fig:testplot_peak1}\,--\,\ref{fig:testplot_peak2} mark the 14\,--\,24\,MeV peak intensities estimated with the model by \citet{ref:RICHARDSON2018}, which are generally consistent 
with the reported results.



While the model is obviously unable to account for the complex event-to-event variations involving, for instance, effects related to pre-existing conditions both in the corona and the interplanetary medium, the calculated spectra are in a reasonable agreement with the observational data within model uncertainties, in terms of both spectral shapes and absolute intensities. 
The largest differences are reported for the 10 September 2017 SEP event at STEREO-A, for which measured intensities are underestimated by more than two orders of magnitude. 
\citet{ref:BRUNO2019} showed that the peak intensities registered by STEREO-A at most energies were associated with the passage of a co-rotating interaction region. In addition, given the $\approx$154$^\circ$ longitudinal deviation between the spacecraft footpoint location and the CME direction, it can be speculated that this discrepancy may be ascribed to an exceptionally extended SEP source -- compared to the $\lesssim$40$^\circ$ $\bar{\sigma}_{\mathrm{sep}}$ value assumed by the model -- in combination with significant transport effects such as cross-field diffusion and IMF co-rotation (see \citealp{ref:BRUNO2019}).
In general, the predicted spectra tend to be softer than observed at the lower end of the energy range, but this is due to the relatively simple spectral form assumed, which does not account for the spectral break that is often observed at a few or tens of MeV (see, e.g., \citealp{ref:ZHAO2016} and references therein). Another limitation is related to the fact that model parameters have been derived by using widespread (three-spacecraft) SEP events, which are usually the most intense. Although not demonstrated by the events in Figures \ref{fig:testplot_peak1}\,--\,\ref{fig:testplot_fluence2}, the model, similar to that of \citet{ref:RICHARDSON2018}, is likely to overestimate the intensity of events associated with relatively slow ($\lesssim$600\,km s$^{-1}$) CMEs.

\section{Summary and Conclusions}\label{Summary and Conclusions}
We presented a new empirical model for predicting SEP event-integrated and peak spectra that is derived from a parameterization of SEP events at 1\,AU based on multi-point spacecraft observations from the twin STEREOs and near-Earth assets, including GOES-13/15 instruments and the PAMELA experiment.
In particular, the data sample used includes 32 SEP events occurring between 2010 and 2014, with a statistically significant proton signal at energies in excess of a few tens of MeV, unambiguously recorded at three spacecraft locations.
The SEP-event spatial distributions in a relatively wide energy range (10\,--\,130\,MeV) have been reconstructed by using a 2D Gaussian functional form accounting for particle-intensity dependence on both heliospheric longitude and latitude. 

This study has also summarized the spatial distributions of these SEP events as a function of 
energy, and further investigation is required to understand
the conditions that give rise to the event-to-event variations. On average, the standard deviation $\sigma_{\mathrm{sep}}$ decreases with increasing energy, from $\approx$39$^\circ$ at 10\,MeV to $\approx$38$^\circ$ at 130\,MeV for the peak intensities, and from $\approx$41$^\circ$ to $\approx$36$^\circ$ for the event-integrated intensities, consistent with the widths found in previous studies. As discussed by \citet{ref:COHEN2017}, this trend can be interpreted in terms of several acceleration and/or transport processes. For instance, lower-energy particles are believed to be efficiently accelerated over a more extended shock region and for a longer time as the shock propagates in the interplanetary space, and they experience more field-line co-rotation, especially during long-duration events, resulting in a broader angular range of intersected field lines.
However, the near energy-independence of the corresponding $\sigma_{\mathrm{sep}}$ value suggests that these effects are less important for 
peak intensities, and that further factors may determine the width. 
For example, as mentioned in Section \ref{SEP spectral analysis}, measured event-integrated intensities are more affected by the contribution from local shock-associated particles and overlapping SEP events, which is expected to be larger at lower energies. In addition, the average distribution peak has been found to be located on field lines with footpoints to the West of the associated CME direction, with the longitudinal deviation increasing with decreasing particle energy. This effect might be accounted for, at least in part, if the high-energy particles were released when the shock is still close to the Sun and hence have peak intensity on field lines with footpoints close to the CME longitude (see \citealp{ref:LARIO2014}). The low-energy distribution may be dominated by particles accelerated by the shock as it moves out through the solar wind, with the largest particle intensities close to the nose of the shock. The nose then intercepts field lines that have footpoints lying to the West of the CME longitude. Figures 10 and 11 of \citet{ref:CANE1988} also demonstrate a similar trend in proton peak intensities with the longitude of the solar event over a range of energies (1\,--\,327\,MeV) for a large sample of events: at low energies the intensity peaks tend to be associated with the passage of shocks following solar events in a broad region around central meridian, whereas at high energies peak intensities are generally associated with well-connected western-hemisphere events and unrelated to shock passage.

We also explored the well-known correlation of particle intensities with the parent CME speed, used as a proxy of the shock acceleration efficiency,
deriving an energy-dependent parameterization.
The results obtained for the analyzed sample of events were used to create a novel empirical model predicting both SEP event-integrated and peak spectra at a given 1\,AU location, that is a development of the SEPSTER model described by \citet{ref:RICHARDSON2018}. The input parameters are given by the velocity of the CME associated with the event [$V_{\mathrm{cme}}$] and by the connection angle with respect to its direction, based on the DONKI catalog, accounting for
the heliographic coordinates ($\alpha_{\mathrm{sc}}$, $\beta_{\mathrm{sc}}$) of the IMF line linking the source to the observer. 
The model calculation is based on the following steps.
\begin{enumerate}
\item The $amplitude$ $\Phi_{\mathrm{o}}(E,V_{\mathrm{cme}})$ can be estimated by using Equations \ref{eq:cme1}-\ref{eq:cme3} and the related parameters reported in Table \ref{tab:cme}. 

\item Then, the intensity at longitude $\beta_{\mathrm{sc}}$ in the 
solar equatorial plane
can be derived by multiplying $\Phi_{\mathrm{o}}(E,V_{\mathrm{cme}})$ by the relative $weight$ $G_{\mathrm{eq}}(E;\beta_{\mathrm{sc}})$ according to Equation \ref{eq:ProjDistr}, with $\alpha_{\mathrm{sep}}$ = $\alpha_{\mathrm{cme}}$; $\beta_{\mathrm{sep}}=\bar{\beta}_{\mathrm{sep}}$ and $\sigma_{\mathrm{sep}}=\bar{\sigma}_{\mathrm{sep}}$ are calculated from Equations \ref{eq:beta}-\ref{eq:sigma} and the parameters reported in Table \ref{spatialpar_table}.

\item To get the final intensity $\Phi(E,V_{\mathrm{cme}};\beta_{\mathrm{sc}})$ at spacecraft location, $\Phi_{\mathrm{eq}}(E;\beta_{\mathrm{sc}})$ 
has to be divided
by the $K_{\alpha}$ factor given by Equation \ref{footlat_corr}, accounting for the spacecraft magnetic footpoint latitude. 

\end{enumerate}
Prediction uncertainties have been estimated by accounting for uncertainties in the various parameters involved.
The model 
was developed using widespread SEP events associated with $>$600\,km s$^{-1}$ fast CMEs in the DONKI catalog, and for proton energies between 10 and 130\,MeV, and it is likely to provide the most reliable calculations for similar events.
The spectra predicted by the model have been tested
by comparing them with the observed spectra for 20 SEP events that were not used to develop the model. A remarkable agreement is found within the model uncertainties, both in terms of the spectral shapes and absolute values, 
despite the many potential 
factors that are neglected in the model but which may influence the particle intensities and cause large event-to-event-event variations.

One practical issue with the model is related to its reliance on the availability and accuracy of near real-time spacecraft CME measurements. The CME may only just have entered the field of view of a space-based coronagraph by the time energetic particles reach a well-connected spacecraft. However, other schemes such as the Relativistic Electron Alert System for Exploration (REleASE: \citealp{ref:POSNER2007,ref:POSNER2009}), based on the early arrival of near-relativistic electrons relative to protons, could be used for predicting the onset of such events; the present model is more suitable for making predictions at less well-connected locations or for long-duration events. Furthermore, at present, space-based coronagraph observations are provided by scientific spacecraft such as SOHO and STEREO and are received with delays of hours or even days after detection; subsequent analysis to obtain CME parameters introduces additional delays. Real-time coronagraph measurements such as will be provided by the NOAA \textit{Space Weather Follow On-Lagrange 1} (SWFO-L1) spacecraft will be required to increase the value of CME-driven SEP prediction schemes, combined with an automated CME identification and tracking system. Another aspect is the accuracy of the CME parameter estimates. As noted by \citet{ref:RICHARDSON2015}, current CME catalogs using different analysis methods generally disagree on the speeds and widths of individual CMEs associated with SEP events. Also, the availability of observations from more than one viewpoint can improve estimates of the CME parameters. For the events considered here, the DONKI CME parameters were generally based on observations from SOHO and the two STEREOs. However, with the loss of STEREO-B, and with STEREO-A approaching Earth at the time of writing, the range of viewpoints is more limited, increasing the uncertainty in the CME parameters.

As discussed by \citet{ref:RICHARDSON2018}, a major problem with a CME-driven prediction scheme is that only a small fraction of CMEs produce an SEP event that is detected at 1\,AU so that the majority of predictions are false. They discuss several ways of increasing the prediction skill of such models by, for example, considering whether Type II or Type III radio emissions accompany the CME, or by just making predictions for those CMEs that exceed a threshold speed or speed and width. This aspect is not considered here, since all of the CMEs in this study were associated with SEP events, but similar considerations would apply to the current model. In addition, ground-based coronagraph observations might be used to provide an earlier warning of a potential SEP event \citep{ref:STCYR2017}.
We also plan to explore the possibility of including in the model parameters related to the related flares, or additional features characterizing the SEP events, such as their onset/peak times or their delay with respect to possibly associated electron events.

%
 \begin{acks}
The DONKI catalog (\url{ccmc.gsfc.nasa.gov/donki/}) is compiled at the CCMC.
The GOES, PAMELA and STEREO data are available at 
\url{www.ngdc.noaa.gov/stp/satellite/goes/}, 
\url{www.ssdc.asi.it/pamela/} and
\url{www.srl.caltech.edu/STEREO/}, respectively. 
The authors thank M.~L. Mays for the assistance with the DONKI database.
They acknowledge support from the NASA/HSR program NNH19ZDA001N-HSR,
the Goddard Space Flight Center / Internal Scientist Funding Model (ISFM) grant HISFM18,
and from the Johnson Space Center / Space Radiation Analysis Group (SRAG) under the Integrated Solar Energetic Proton Alert/Warning System (ISEP) project. 
I.G. Richardson also acknowledges support from NASA program NNH17ZDA001N-LWS and from the STEREO mission.
 \end{acks}
 
 {\footnotesize\paragraph*{Disclosure of Potential Conflicts of Interest}
The authors declare that they have no conflicts of interest.}
 
\appendix   
\section{Estimate of the Magnetic Footpoint Location}\label{Appendix A}
The HGS coordinates ($\alpha_{\mathrm{sc}}$,$\beta_{\mathrm{sc}}$) of the footpoint location of the IMF line passing through a given spacecraft can be estimated by assuming a simple 3D Parker spiral model \citep{ref:PARKER1958}. Specifically, the longitude is calculated as:
\begin{equation}
\beta_{\mathrm{sc}} \approx b_{\mathrm{sc}} + \frac{\Omega_{\odot} }{V_{\mathrm{sw}}}\hspace{1mm} \left( R_{\mathrm{sc}} - R_{0}' \right) \hspace{1mm} \cos(\alpha_{\mathrm{sc}}),
\end{equation}
where $b_{\mathrm{sc}}$ is the longitude of the spacecraft location,
$R_{\mathrm{sc}}$ is its radial distance, 
\begin{equation}
R_{0}' = R_{0} \hspace{1mm} \left[ 1 + \log\left( \frac{R_{\mathrm{sc}}}{R_{0}} \right) \right],
\end{equation}
with $R_{0}$=2.5\,R$_{\odot}$ the radius of the ``source'' surface;
$V_{\mathrm{sw}}$ is the solar-wind speed [km s$^{-1}$] -- assumed to be constant and purely radial --
and $\Omega_{\odot}$ is the differential solar rotation rate at the latitude $\alpha_{\mathrm{sc}}$ of the footpoint location:
\begin{equation}
\Omega_{\odot} = A - B \hspace{1mm}\sin^{2}(\alpha_{\mathrm{sc}}) - C\hspace{1mm}\sin^{4}(\alpha_{\mathrm{sc}}),
\end{equation}
with $A$=2.972$\pm$0.010\,$\mu$rad s$^{-1}$, $B$=0.484$\pm$0.038\,$\mu$rad s$^{-1}$ and $C$=0.361$\pm$0.051\,$\mu$rad s$^{-1}$ \citep{ref:SNODGRASS1990};
in particular, the value of $A$ is related to the sidereal rotation period at the equator ($\approx$24.47 days).
The footpoint latitude $\alpha_{\mathrm{sc}}$ is assumed to coincide with the heliographic latitude of the central point of the solar disk as seen by the spacecraft, 
accounting for the Sun's rotation axis' tilt of about 7.25$^\circ$ relative to the Ecliptic plane. 
In general, at large radial distances the Parker spiral IMF is simplified as an Archimedes spiral by neglecting the logarithmic term ($R_{\mathrm{sc}}-R_{0}'\approx R_{\mathrm{sc}}-R_{0}$).

The uncertainty associated with the footpoint longitude can be calculated as:
\begin{equation}
\delta\beta_{\mathrm{sc}} = \sqrt{(\delta\beta_{\mathrm{sw}})^{2}+(\delta\beta_{\mathrm{tra}})^{2}},
\end{equation}
where
\begin{equation}
\delta\beta_{\mathrm{sw}} = \frac{(R_{\mathrm{sc}} - R_{0}')\hspace{1mm}\cos(\alpha_{\mathrm{sc}})}{V_{\mathrm{sw}}} \hspace{1mm}\sqrt{ \left[ \left( \frac{\partial\Omega_{\odot}}{\partial\alpha_{\mathrm{sc}}} - \tan(\alpha_{\mathrm{sc}})\hspace{1mm}\Omega_{\odot}\right) \delta\alpha_{\mathrm{sc}} \right]^{2} + \left[ \frac{\Omega_{\odot}}{V_{\mathrm{sw}}}\hspace{1mm}\delta V_{\mathrm{sw}} \right]^{2}},
\end{equation}
with $\delta V_{\mathrm{sw}}$ the uncertainty in the solar-wind speed estimate,
and $\delta\beta_{\mathrm{tra}}$ accounts for interplanetary transport effects ignored by the model.
Similarly, the error on the footpoint latitude takes into account the deviations from the spiral-model approximation out of the equatorial plane and effects related to
the differential changes across the solar surface.

\bibliographystyle{spr-mp-sola}
\bibliography{bibliography}
%
%
%
%
%

\end{article} 
\end{document}